\documentclass[preprint,12pt]{elsarticle}
\usepackage{amssymb}
\journal{Nuclear Physics A}

\newcommand{\ba}{\begin{eqnarray}}
\newcommand{\ea}{\end{eqnarray}}
\newcommand{\beqs}{\begin{eqnarray}}
\newcommand{\eeqs}{\end{eqnarray}}

\begin{document}

\begin{frontmatter}

\title{Total cross sections and $\rho$ at high energy}

\author{
O. V. Selyugin %\fnmsep\thanks{\email{selugin@theor.jinr.ru}  }
}
%\address{}
\address{ BLTPh, JINR, Dubna, Russia}

\begin{abstract}
  Analysis of the new experimental data obtained by the TOTEM Collaboration at LHC at $\sqrt{s} = 7 \ $TeV at small momentum transfer  is presented. The impact of the different assumptions on the extraction of the parameters of the elastic scattering amplitude, especially on the  size of the total cross sections, is examined. It is shown that the contribution of the Coulomb amplitude and Coulomb-hadron interference term should be taken into account in the analysis of the existing experimental data at small momentum transfer. Our new method of  extracting the real part of the hadron scattering amplitude from  experimental data shows the inconsistency of  the size of $\rho=0.14$ to the parameters of the imaginary part of the hadron scattering amplitude obtained by the TOTEM Collaboration. The analysis of the data is  compared with the similar  analysis  in the Regge approach for the hadron scattering amplitude.
\end{abstract}

\begin{keyword}
Hadron, total cross sections,  elastic scattering, high energies
% MSC codes here, in the form: \MSC code \sep code
% or \MSC[2008] code \sep code (2000 is the default)
\end{keyword}

\end{frontmatter}

%\PACS{
%      {13.40.Gp}, %{discribing text of that key} \and
%      {14.20.Dh}, %{text} % \and
%      {12.38.Lg} %{ text} }

%\section{Introduction}

\section{Introduction}

    The determination of the parameters of the elastic scattering amplitude,
     such as  $\sigma_{tot}$ - total cross sections, $\rho(s,t)$ - ratio of the real to imaginary part
     of the scattering amplitude, $B(s,t)$ - slopes of the imaginary and real parts of the scattering amplitude
     is  one of the important tasks of  experimental researches at the LHC.
     The properties of the elastic scattering amplitude at small angles are tightly connected,
     on the one hand, with the first principles of the theory of  strong interactions and, on the other hand, with the non-perturbative
     properties of the hadron interaction. The knowledge of the size of the total, elastic and inelastic cross sections is
      important for other experimental researches at LHC.
As part of the SMC Collaboration of the LHC CERN the   TOTEM Collaboration (TOTal and Elastic scattering cross-section Measurement) is  one of the special purpose experiments
     to obtain the new information about the elastic hadron scattering at LHC energies
     at wide momentum transfer \cite{TOTEM0}.  Under various beam and background conditions, the differential elastic and elastic, inelastic total proton proton cross sections have been measured. Now they published \cite{TOTEM-11a} the data on the differential elastic cross sections
     at $\sqrt{s} = 7 $ TeV
     at the sufficiently large momentum transfer
     ($0.3 \leq |t| \leq 2.5 $ GeV$^2$ and small $5 \cdot  10^{-3} \leq |t| \leq 0.4 $ GeV$^2$
     using the size of the Luminocity obtained by the SMC Collaboration.
      They obtained that the slope $B= 19.9 \pm 0.3 $ GeV$^{-2}$
     under the assumption that the differential cross section in the region of
     $5 \cdot 10^{-3} \leq |t| \leq 0.2 $ GeV$^2$ can be described by exponent
     with the slope independent of $t$.
     Extrapolation of the differential cross sections at $t=0$ and the optical theorem
     gave the size of the total cross sections $\sigma_{tot}= 98.7 $ mb.
     They used the size of  $\rho=0.141 \pm 0.007$
     (ratio of the real to imaginary part of the elastic scattering amplitude)
      from the analysis carried out by the COMPETE
     (Computerised Models, Parameter Evaluation for Theory and Experiment) Collaboration
     \cite{COMPETE}  as preferred-model extrapolation.
 The COMPETE Collaboration considered several hundreds of possible parametrization
 for $pp$, $\bar p  p$, $\pi^\pm p$, $K^\pm p$, $\gamma p$ and $\gamma \gamma$,
 based on simple, double
or triple poles, and kept only those which had a global $\chi^2$/point smaller than 1, for
$\sqrt{s}\geq 5$ GeV. From these, one can predict $\rho$ and $\sigma_{tot}$
% the total cross section
at the LHC, and
estimate the error due to the extrapolation.
     The first results obtained by the TOTEM collaboration
     \cite{TOTEM-11a,TOTEM-11} on the differential cross sections are in disfavor with   practically all theoretical model predictions  \cite{TOTEM-11a,Rev-LHC}.
     Now the models are reconstructed to obtain the coincidence with the new experimental results.

The number of elastic events is related to the total hadronic cross
section through
% the following formula:
\begin{eqnarray}
\frac{dN}{dt}=&&{\cal{L}} \left[\frac{4 \pi \alpha^2}{|t|^2} G^4(t)\right.
%\nonumber \\
  - \frac{2 \alpha\left(\rho(s,t) + \phi_{CN}(s,t)\right) \sigma_{tot}G^2(t) e^{-\frac{B(s,t)|t|}
{2}}}{|t| }
      \nonumber \\
  &&   \left. +\frac{\sigma_{tot}^2(1+\rho(s,t)^2)e^{-B(s,t)|t|}}{16 \pi}\right]
\label{dN/dt}
 \end{eqnarray}
 where the three terms are due to the Coulomb scattering,
Coulomb-hadron interference and hadronic interactions;
${\cal{L}}$  is the integrated luminosity, $\alpha$ is the electromagnetic
coupling constant, $\phi_{CN}(s,t)$ is the Coulomb-hadron phase, and $G(t)$ is the
electromagnetic form factor given by
 \begin{eqnarray}
G(t) = \frac{4 m_p^2 - \mu t }{4 m_p^2-t}\frac{\Lambda^2}{(\Lambda - t)^2}.
\label{emff}
 \end{eqnarray}
  with $m_p$ being the proton mass, $\Lambda=0.71$ GeV$^2$ and $\mu=2.79$.

          When we extract the parameters of the scattering amplitude from the experimental data,
          we need to use some theoretical assumption and approximations. For example,
       in \cite{dif04,CS-PRL09} it was shown that the saturation regime, which can  occur at the LHC energies,
       changes the behavior of the slope of the differential cross sections at small momentum transfer.
       As a result, the differential cross section cannot be described by  a simple exponential form with
       the constant slope.  Note, in (\ref{dN/dt}) the assumption of the equality of the slopes of the imaginary
       and real parts  ( $B_{Re}(s,t)=B_{Im}(s,t) $ )    was used.

        A  remarkable example was obtained from the analysis of the experimental data at $ Sp\bar{p}S$.
       In the proton-antiproton scattering at $ \sqrt{s}=540 \ $GeV there are two different measures of the
       size of $\rho$ : the UA4 Collaboration ($\rho=0.24$) and the UA4/2  Collaboration ($\rho=0.139$). However,
       more careful analysis gave $\rho=0.19$ for the data of UA4 \cite{Sel-YF92} and  $\rho=0.16$  for the data
       UA4/2  \cite{Sel-UA42}). Hence, the contradictions between the experimental data practically
       disappeared.
       In \cite{Sel-UA42}, it was shown that  $\chi^2$ in the fitting of the experimental data decreases
       by $10 \% $ if we use the slope in a more complicated form $B(t)=B_0 t + C \sqrt{t_{0}-t}$.
       It changes the form of the differential cross section at very small momentum transfer.
  %     In this case, the size
  %     of  $\rho$ increases up to $0.16 \div 0.17$.

                In the present paper, we made some analysis of the new experimental data obtained by the
                TOTEM Collaboration at $\sqrt{s} = 7 \ $TeV at small momentum transfer.
%
    %          First,
              We used the phenomenological model with exponential behavior of the
              scattering amplitude.
%              Sometimes they think that is non-model approaches. Really,
              It is the simplest model with many different assumptions. Some of them will be
              discussed in this paper. In this model, we need to take  $\rho(s,t)$ as a free parameter
              or, as made by the TOTEM Collaboration, from the previous fit of the COMPETE Collaboration.
              Then we take the simple Regge approaches, where the real part of the scattering amplitude
              is determined by the intercept $\alpha(0)-1$ of the  Regge trajectory.
              % size of the intercept.

  The differential cross
  sections of the nucleon-nucleon elastic scattering  can be written as the sum of different
  helicity  amplitudes:
\begin{eqnarray}
  \frac{d\sigma}{dt} =
 \frac{2 \pi}{s^{2}} (|\Phi_{1}|^{2} +|\Phi_{2}|^{2} +|\Phi_{3}|^{2}
  +|\Phi_{4}|^{2}
  +4 | \Phi_{5}|^{2} ). \label{dsdt}
\end{eqnarray}
\renewcommand{\bottomfraction}{0.7}
  The total helicity amplitudes can be written as a sum of nuclear $\Phi^{h}_{i}(s,t) $  and electromagnetic $\Phi^{e}_{i}(s,t) $
  amplitudes
   $\Phi_{i}(s,t) =
  F^{h}_{i}(s,t)+F^{\rm em}_{i}(s,t) e^{\varphi(s,t)} $\,,
%  where  $F^{h}_{i}(s,t) $ comes from the strong interactions,
% $F^{\rm em}_{i}(s,t) $ from the electromagnetic interactions and
where  $\varphi(s,t) $
 is the interference phase factor between the electromagnetic and strong
 interactions \cite{selmp1,selmp2,Selphase}.
 We assume, as usual, that at high energies and small angles the one-flip and double-flip hadron amplitudes are small with respect to the spin-nonflip ones and that the hadron contributions to $\Phi_1$ and $\Phi_3$ are the same.
 %, as are the electromagnetic ones.
% For the hadron part the amplitude with spin-flip is neglected in this approximation, as usual at high energy.
 % The electromagnetic amplitude can be calculated in the framework of QED.

 The electromagnetic amplitude can be calculated in the framework of QED.
%  and in the one-photon approximation    they are \cite{bgl} %   , we have \cite{bgl}
 % In the high energy approximation, it can be  obtain \cite{bgl}:
    In the high energy approximation, it can be  obtained \cite{bgl}
  for the spin-non-flip amplitudes:
  \begin{eqnarray}
  F^{em}_{1}(t) = \alpha f_{1}^{2}(t) \frac{s-2 m^2}{t}; \ \ \  F^{em}_3(t) = F^{em}_1;
%   \\ \nonumber
  \end{eqnarray}
   and for the spin-flip amplitudes:
 \begin{eqnarray}
  F^{em}_2(t) =  \alpha  \frac{f_{2}^{2}(t)}{4 m^2} s; \ \ \
%   F^{em}_3 &=& F^{em}_1, \\
   F^{em}_{4}(t) =  - F^{em}_{2}(t), \ \  \  \\ \nonumber
  F^{em}_5(t) =  \alpha \frac{s }{2m \sqrt{|t|}} f_{1}(t) \ f_{2}(t),
  \end{eqnarray}
%  For numerical calculations let us take the hadron amplitude in the form
 % (\ref{fh}).
  where the form factors are:
    \begin{eqnarray}
    f_{1}(t) = \frac{4 m_{p}^{2} - \mu \ t}{ 4 m_{p}^{2} - \ t} \ G_{d}(t); \ \ \     % \\ \nonumber
    f_{2}(t) = \frac{4 m_{p}^{2} \ (\mu-1)}{ 4 m_{p}^{2} - \ t} \ G_{d}(t);
    \label{emff}
\end{eqnarray}
  where
  %$k$ is relative to the anomalous magnetic moment, and
  $G_{d}(t)$ has the conventual dipole form
 \begin{eqnarray}
   G_{d}(t)= 1/(1-t/0.71)^2.
  \end{eqnarray}

\section{Experiment and analysis}
%\section{Analysis of the differential cross sections}
%  \vspace{10cm}
    The hadron spin non-flip amplitude was chosen in the form
    taking into account the possible non-exponential form
  (using the nearest $t$-channel singularity \cite{q1,q3,Rev-LHC})
    \begin{eqnarray}
 F(s,t)= (i+\rho)\frac{\sigma_{tot}}{4 k \pi } e^{[B/2 \ t \ +C /2 \ (\sqrt{4m_{\pi}^2-t}-2m_{\pi})] ) }
 %\ exp(B/2 \ t \ -C/2 \  q)
 \label{fh}
  \end{eqnarray}
 where $m_{\pi}=0.139$ GeV is pion mass and $k=0.38938 \ $ mb GeV$^{-2}$,  $t=-q^2$, and $C$ GeV$^{-1}$ is some coefficient which is determined by some additional part of the slope.
  %In the case
 % of $C \neq 0$,  the differential cross sections can grow faster or   slowly as $t \rightarrow 0$. % at small momentum transfer.
 In most part we will examine the set of the TOTEM data at small $t$ with   $N=47$  points and $-t_{max} \ = \ 0.112$ GeV$^2$
 and some shortest interval with  $  N=40 $  points and $-t_{max} \ = \ 0.097$ GeV$^2$.
     The whole set ($  N=86 $  points and $-t_{max} \ = \ 0.3 \ $GeV$^2$ we will examined only as an example.
     This interval of momentum transfer is large and the imaginary part of the scattering amplitude may have some
     complicated form.
      In all our calculations we used only statistical errors of the experimental data.

% Table 1a
      \begin{table}[h]
\label{tab:1}       % Give a unique label
\begin{center}
\begin{tabular}{c|c|c|c|c|c|c}
\hline\noalign{\smallskip}
 i & $N$  & $\sum_{i=1}^{N} \chi^{2}_{i}$ &$\rho$  & $B$ & $C$ & $\sigma_{tot}, mb$  \\\hline
   & & & & & &\\
 1 & 86 & 287. & $0.14 $ fix  & $20. $ & $0.fix $ & $ 98.87\pm0.1  $   \\
 2 & 86 &   287 & $ 0.05 fix $& $20.$  & $ 0.fix          $  &  $   99.7\pm0.1  $    \\
 3 & 86 &   287 & $ 0.146\pm0.3  $& $20.$  & $ 0.fix          $  &  $   98.8\pm0.4  $    \\
 4 & 86 &   220.5 & $  0.14 fix       $& $21.7$  & $-1.4\pm0.2    $  &  $  97.9\pm0.2   $     \\
 5 & 86 &   220. & $  0.05 \pm 0.4       $& $21.8$  & $ -1.4\pm0.2    $  &  $  98.76\pm4.   $     \\
\noalign{\smallskip}\hline
\end{tabular}
\caption{The basic parameters of the model are determined by fitting experimental data without the electromagnetic
 contributions and with free $\sigma_{tot}$.}
\end{center}
\end{table}

% Table 2
      \begin{table}[h]
\label{tab:1}       % Give a unique label
\begin{center}
\begin{tabular}{c|c|c|c|c|c|c}
\hline\noalign{\smallskip}
 i & $N$  & $\sum_{i=1}^{N} \chi^{2}_{i}$ &$\rho$  & $B$ & $C$ & $\sigma_{tot}, mb$  \\\hline
   & & & & & &\\
 1 & 47 & 64.96 & $0.176\pm0.2 $  & $19.9 $ & $0.fix $ & $ 98.05\pm1.7  $   \\
 2 & 47 & 64.96 & $0.15 fix$  & $19.9 $ & $0.fix $ & $ 98.47\pm0.1  $   \\
 3 & 47 & 64.96 & $0.14 fix$  & $19.9 $ & $0.fix $ & $ 98.6\pm0.1  $   \\
 4 & 47 & 64.96 & $  0.1fix       $& $19.9  $  & $ 0.fix          $  &  $   99.1\pm0.1   $    \\
 5 & 47 &   64.96 & $ 0.05 fix $& $19.9$  & $ 0.fix          $  &  $   99.44\pm0.1  $    \\
 6 & 47 &   64.96 & $ 0.0 fix $& $19.9$  & $ 0.fix          $  &  $   99.57\pm0.1  $    \\
 7 & 47 &   64.96 & $-0.05 fix $& $19.9$  & $ 0.fix          $  &  $   99.44\pm0.1  $    \\
 8 & 47 &   61.09 & $  0.14 fix       $& $18.5$  & $ 1.05\pm0.54    $  &  $  98.99\pm0.2   $     \\
 9 & 47 &   61.09 & $  0.1 fix       $& $18.5$  & $ 1.06\pm0.54    $  &  $  99.47\pm0.2   $     \\
 10& 47 &  61.09  & $ 0.0 fix  $& $18.5$  & $ 1.07 \pm0.54   $  &  $  99.97\pm0.2   $    \\
 11&  47 &  60.08  & $-0.03 \pm0.1   $& $18.4$  & $ 1.07 \pm0.54   $  &  $  99.94\pm0.4   $    \\
\noalign{\smallskip}\hline
\end{tabular}
\caption{The basic parameters of the model are determined by fitting experimental data without the electromagnetic
 contributions. % and with free $\sigma_{tot}$.
 }
\end{center}
\end{table}

    First, let us make the fit of the differential cross sections with the hadronic amplitude in form (\ref{fh})
    and  not take into account the electromagnetic interactions.
    The result of the fit for $N=86$ is presented in Table 1. The $\sum_{i=1}^{N} \chi^{2}_{i}$ is large
    for all variants.

    Hence, let us examine the interval of $t$ like the interval of the experimental data of the UA4/2 Collaborations. It includes the $N=47$
    experimental points.
    The result of the fit is presented in Table 2.
% Table 3
    \begin{table}[h]
\label{tab:2}       % Give a unique label
\begin{center}
\begin{tabular}{c|c|c|c|c|c}
\hline\noalign{\smallskip}
 $N$  & $\sum_{i=1}^{N} \chi^{2}_{i}$ &$\rho$  & $B$ & $C$ & $\sigma_{tot}, mb$  \\\hline
  & & & & &\\
 86 & 281. & $0.14 $fixed  & $20. $ & $0.fix $ & $ 99.4\pm0.1  $   \\
 86 &   281. & $  0.1fix       $& $20.  $  & $ 0.fix          $  &  $   99.7\pm0.1   $    \\
 86 &   288. & $ 0. fix $& $20.$  & $ 0.fix          $  &  $   99.8\pm0.1  $    \\
 86 &   245. & $  0.14fix       $& $21.3$  & $-1.03 \pm0.2    $  &  $  98.6\pm0.2   $     \\
 86 &  215.  & $ 0.0 fix    $& $21.8$  & $-1.2 \pm0.2   $  &  $  98.7\pm0.2   $    \\
 86 &  175.  & $-0.41 \pm0.1   $& $23.2$  & $-2.77 \pm0.2   $  &  $  89.1\pm3.   $    \\
\noalign{\smallskip}\hline
\end{tabular}
\caption{The basic parameters of the model are determined by fitting experimental data with $N=86$. % free $\sigma_{tot}$.
}
\end{center}
\end{table}
% Table 4
    \begin{table}[h]
\label{tab:2}       % Give a unique label
\begin{center}
\begin{tabular}{c|c|c|c|c|c}
\hline\noalign{\smallskip}
 $N$  & $\sum_{i=1}^{N} \chi^{2}_{i}$ &$\rho$  & $B$ & $C$ & $\sigma_{tot}, mb$  \\\hline
  & & & & &\\
 47 & 87.1 & $0.2 $fixed  & $20.1 $ & $0.fix $ & $ 98.5\pm0.1  $   \\
 47 & 77.1 & $0.14 $fixed  & $20. $ & $0.fix $ & $ 99.2\pm0.1  $   \\
 47 &   71.6 & $  0.1fix       $& $20  $  & $ 0.fix          $  &  $   99.5\pm0.1   $    \\
 47 &   61.1 & $ -0.07\pm0.05 $& $19.8$  & $ 0.fix          $  &  $   98.93\pm0.8  $    \\
 47 &   61.2 & $  0.1fix       $& $17.7$  & $ 1.66 \pm0.54    $  &  $  100.1\pm0.2   $     \\
 47 &  60.6  & $ 0.0 fix    $& $18.8$  & $ 0.82 \pm0.54   $  &  $  99.8\pm0.2   $    \\
 47 &  60.6  & $ 0.01 \pm0.1   $& $18.9$  & $ 0.74 \pm0.8   $  &  $  99.7\pm0.8   $    \\
\noalign{\smallskip}\hline
\end{tabular}
\caption{The basic parameters of the model are determined by fitting experimental data with  $N=47$. % free $\sigma_{tot}$.
}
\end{center}
\end{table}
 The first row shows the calculation with % free
   variation of three parameters:
  $\rho$, slope - $B$ and $\sigma_{tot}$.
        The minimum in $\sum_{i=1}^{N} \chi^{2}_{i}$ in the fitting procedure is very wide
        which leads to the large errors of  the  determined sizes of $\rho$ and $\sigma_{tot}$.
        % are determined with .
        Let us fix the size of $\rho$.
         %and determine the sizes of the slope and $\sigma_{tot}$.
       The next 6  rows (2-7)  present the fit with a different fixed size of $\rho$ and with $C=0$.
       %zero size of the        additional part of the slope.
       We see that $\sum_{i=1}^{N} \chi^{2}_{i}$ is independent of the size of $\rho$ and $\sigma_{tot}$
        has a small change. The size of  $\sigma_{tot}$ is slightly above the data obtained by the TOTEM Collaboration  \cite{TOTEM-11}.
        In the next 3 rows (8-10) the fit includes an additional part of the slope which is proportional to the coefficient $C$.
        The $\chi^2$ decreases slightly, which reflects  the presence of the additional free parameter, but again
        the size of  $\rho$  is badly determined. The size of $\sigma_{tot}$ increases slightly, but the errors of
        $\sigma_{tot}$ increase essentially. The last row (11) presents the attempt to fit with all
        free parameters. In this case, we obtain a small value of $\rho$ and large $\sigma_{tot}$.

%Fig 1
%\begin{figure}[htb]
%%\resizebox{0.49\textwidth}{!}{%
% \includegraphics{totrho1.ps}
%\caption{Size of $\sigma_{tot}$ over $\rho$ in two variants a) (hard line) - without electromagnetic interaction
%and b) (dashed line) with electromagnetic interaction.
%}
%\label{fig:misse}       % Give a unique label
%\end{figure}

   Hence, we can conclude that in neglecting
      the  electromagnetic contribution the size of  $\rho$ % and the form of the slope
   practically does not impact the determination of  $\sigma_{tot}$ in this region of momentum transfer.
    Such assumption was made by the TOTEM Collaboration in the fitting procedure.

         Now let us make the same fit but include the electromagnetic part of the elastic scattering amplitude.
         Of course, it is sufficiently small in this range of the momentum transfer; however, it leads to
         visible results.
         The fits for the number of the experimental points $86$
         are shown in Table 3.
         The size of $\sum_{i=1}^{N} \chi^{2}_{i}$ is large
         except for the last  row. But in this case we obtain the unusual sizes of $\rho$
    and $\sigma_{tot}$.

  \begin{table}[h]
\label{tab:5}       % Give a unique label
\begin{center}
\begin{tabular}{c|c|c|c|c|c}
\hline\noalign{\smallskip}
 $N$  & $\sum_{i=1}^{N} \chi^{2}_{i}$ &$\rho$  & $B$ & $C$ & $\sigma_{tot}, mb$  \\\hline
  & & & & &\\
  40& 78.8 & $0.2 fix $ & $20.1 $ & $0. fix $ & $ 98.6 \pm0.12 $   \\
  40& 70.4 & $0.14 fix $ & $20. $ & $0. fix $ & $ 99.29 $   \\
 40 &   65.8 & $  0.1 fix       $& $20  $  & $ 0. fix          $  &  $   99.55 \pm0.12  $    \\
 40 &   56.6 & $ -0.076\pm0.06 $& $19.8$  & $ 0. fix          $  &  $   98.83  \pm0.12 $    \\
 40 &   54.7 & $  0.1 fix       $& $16.3$  & $ 2.63\pm0.8   $  &  $  100.3  \pm0.27 $     \\
 40 &   54.9 & $  0.0 fix      $& $17.8$  & $ 1.47\pm0.8 $  &  $  99.96 \pm0.26  $     \\
 40 &  54.6  & $ 0.06\pm0.01  $& $16.9$  & $ 2.17 \pm1.4   $  &  $  100.3  \pm0.3 $    \\
\noalign{\smallskip}\hline
\end{tabular}
\caption{The basic parameters of the model are determined by fitting experimental data with $N=40$. % free $\sigma_{tot}$.
}
\end{center}
\end{table}

  \begin{table}[h]
\label{tab:6}       % Give a unique label
\begin{center}
\begin{tabular}{c|c|c|c|c|c}
\hline\noalign{\smallskip}
 $N$  & $\sum_{i=1}^{N} \chi^{2}_{i}$ &$\rho$  & $B$ & $C$ & $\sigma_{tot}, mb$  \\\hline
  & & & & &\\
 47 &134.5 & $0.14 $fixed  & $19.8$ & $0. - fixed $ & $ 98.4  $   \\
 47 &  174.7 & $  0.1fix       $& $19.7$  & $ 0.fix          $  &  $   98.4       $    \\
 47 &   88.1 & $ 0.203\pm0.01 $& $20.1 $  & $ 0.$fixed          &  $   98.4       $     \\
 47 &  105.3 & $  0.14 $fixed& $22.9$  & $-2.3\pm 0.3$fixed  &  $   98.4       $    \\
 47 &  61.4  & $-0.105\pm0.02  $& $20.$  & $-0.14 \pm0.4   $  &  $   98.4       $    \\
\noalign{\smallskip}\hline
\end{tabular}
\caption{The basic parameters of the model are determined by fitting experimental data with fixed $\sigma_{tot}$.}
\end{center}
\end{table}

    \begin{table}[h]
\label{tab:7}       % Give a unique label
\begin{center}
\begin{tabular}{c|c|c|c|c|c}
\hline\noalign{\smallskip}
 $N$  & $\sum_{i=1}^{N} \chi^{2}_{i}$ &$\rho$  & $B$ & $C$ & $\sigma_{tot}, mb$  \\\hline
  & & & & &\\
 40 &   88.17 & $  0.203\pm0.007  $   & $20.1 $  & $ 0. - fixed     $  &  $   98.4   $   \\
 40 &   88.2 & $  0.2  $fixed   & $20. $  & $ 0. - fixed     $  &  $   98.4   $   \\
 40 &  125.6 & $  0.14 $fixed   & $19.7$  & $ 0. - fixed     $  &  $   98.4   $   \\
 40 &  157.7 & $  0.1fix       $& $19.6$  & $ 0.fix          $  &  $   98.4   $    \\
 40 &  102.4 & $  0.14fix      $& $22.4$  & $-1.75 \pm0.4   $  &  $   98.4   $     \\
 40 &   56.9 & $ -0.106\pm0.01 $& $19.8$  & $ 0.fix          $  &  $   98.4   $    \\
 40 &  56.8  & $-0.11\pm0.02  $& $19.7$  & $ 0.1 \pm0.7  $  &  $   98.4   $    \\
\noalign{\smallskip}\hline
\end{tabular}
\caption{The basic parameters of the model are determined by fitting experimental data with fixed $\sigma_{tot}$.}
\end{center}
\end{table}

            In the case with $N=47$ (Table 4), $\sum_{i=1}^{N} \chi^{2}_{i}$ is larger than  in Table 2 for the first
            three rows, but for other cases it is almost the same.
          The first 4 rows present the fit without the additional part of the
         slope.  In this case, the influence of the size of $\rho$ is visible and we can make the fit taking
         $\rho$ as free parameters (the row 4). We obtained the negative size of $\rho$
         at   minimum  $\sum_{i=1}^{N} \chi^{2}_{i}$.
         It is essentially far away from the TOTEM Collaboration analysis and the predictions of the COMPETE Collaboration.
         However, the size of $\sigma_{tot}$ is the same as in  Table 2 in the region of  errors.
         If we take the additional part of the slope (the rows 5-7 of  Table 4),
          the error of $\rho$ increases and its size is badly determined.
         But the size of the coefficient $C$ is determined well especially with the fixed size of $\rho$.
         In this case, the size of $\sigma_{tot}$ increases.
         %  with middle   value $99.7 \ $mb but,
         % of course, with a large error.

         If a slightly less interval of momentum transfer ($40 \ $points with $-t \leq 0.097 \ $GeV$^2$) is taken,
         the whole picture will be the same (see Table 5),  but the size of the coefficient $C$ is determined better and
         its size reaches $2.6 \pm 0.8 \ $GeV$^{-1}$. Remarkable, the size of  $\sigma_{tot}$ is
         practically the same as in the previous case.

     We can check some parts of our assumptions included in the fit procedure. If we know the parameters of
     the imaginary part of the hadron scattering amplitude, then it is possible to calculate the real part \cite{Sel-YF92}
     using the  experimental data on the differential cross sections
     and taken into account that for the proton-proton interaction $Re F_{C(}(t) < 0$.
  \begin{eqnarray}
  ReF^{h} (t_{i})=&& - Re F_{C}(t_{i}) \\ \nonumber
   \pm &&[[\frac{d\sigma}{dt_{i}}|_{exp.} - k \pi *(Im F_{C}(t_{i}) + ImF_h)^2(t_{i})]/(k \pi)]^{1/2}, \label{refh3}
%   \\ \nonumber
  \end{eqnarray}
here  the sign $(+)$ in the case $ReF_{h} \ge |ReF_{C}|$ and sign $(-)$ in
    the case when  $|ReF_{C}| > ReF_{h}$.
   Let us take  the imaginary part of the hadron scattering amplitude  in the simple exponential form
      with the parameters obtained by the  TOTEM Collaboration
     \begin{eqnarray}
  ImF^{h} (t)=  \sigma_{tot}/(4 k \pi ) e^{B t /2} ,
  \label{im}
%   \\ \nonumber
  \end{eqnarray}
  with $\sigma_{tot}=98.6  \ $mb and $B=19.9 \ $GeV$^{-2}$.

    The obtained results are shown in Fig.1 by the triangles (when we neglect the Coulomb amplitude) and
    by the squares (with taking into account the Coulomb amplitude. If the square of the sum of the imaginary parts
    of the Coulomb and hadron amplitudes exceeds the experimental data,
    then the value under the square root will be negative and
     the imaginary part of the calculations appears.
      With the imaginary part of the amplitude calculated with the parameters of the TOTEM Collaboration we obtain the same situation
     for many experimental points at larger momentum transfer in both cases (with and without the electromagnetic contributions).
      In Fig.1, such an imaginary  part (multiplied by $-i$) is presented by the empty triangles and squares.
    Compare such results with the real part which is assumed in the fit procedure of the TOTEM Collaboration
    \begin{eqnarray}
  Re F^{h} (t)=  \rho \ \sigma_{tot}/(4  k \pi) e^{B t /2} ,
%   \\ \nonumber
\label{reim}
  \end{eqnarray}
  with $\rho=0.14$ which is obtained by the COMPETE analysis and used by the TOTEM Collaboration.
  In Fig.1, this real part of the hadron amplitude is presented by the long -dashed line.
    Obviously,  the line coincides in most part, except the regions of the very small and large $t$, with the case when we take into account the
    Coulomb amplitude.
      We also see that at large momentum transfer the contribution of the imaginary part of the scattering amplitude exceeds
      the experimental data. This effect shows either  some problem with the normalization,
      or the non-exponential form of the hadron scattering amplitude.

  %==================Figs.1 =============================
\label{sec:figures}
\begin{figure}[htb]
\vspace{-1.5cm}
\centerline{\includegraphics[width=0.7\textwidth]{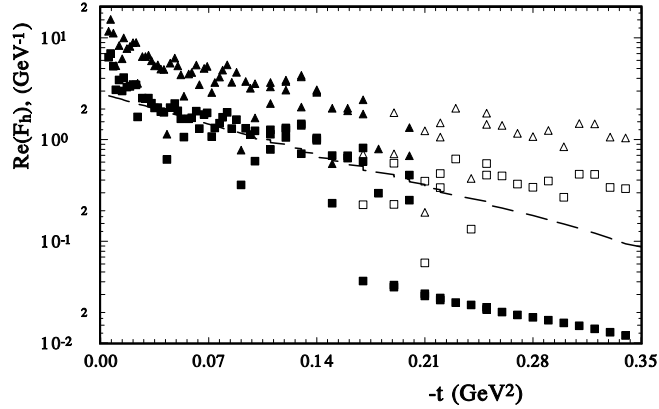}}
\caption{ Real part of the hadronic amplitude calculated by eq.(9)(triangles and squared without and with $F_c$;
[solid and empty represent real and imaginary parts of eq.(9) see text];
 long dashed line - the calculations by eq.(\ref{reim})).
}\label{Fig:1}
\end{figure}
%==================Figs.2 =============================
\label{sec:figures}
\begin{figure}[htb]
\vspace{-1.5cm}
%\centerline{\includegraphics[width=0.35\textwidth]{smith_joe_eds09_fig1.eps}}
\includegraphics[width=0.5\textwidth] {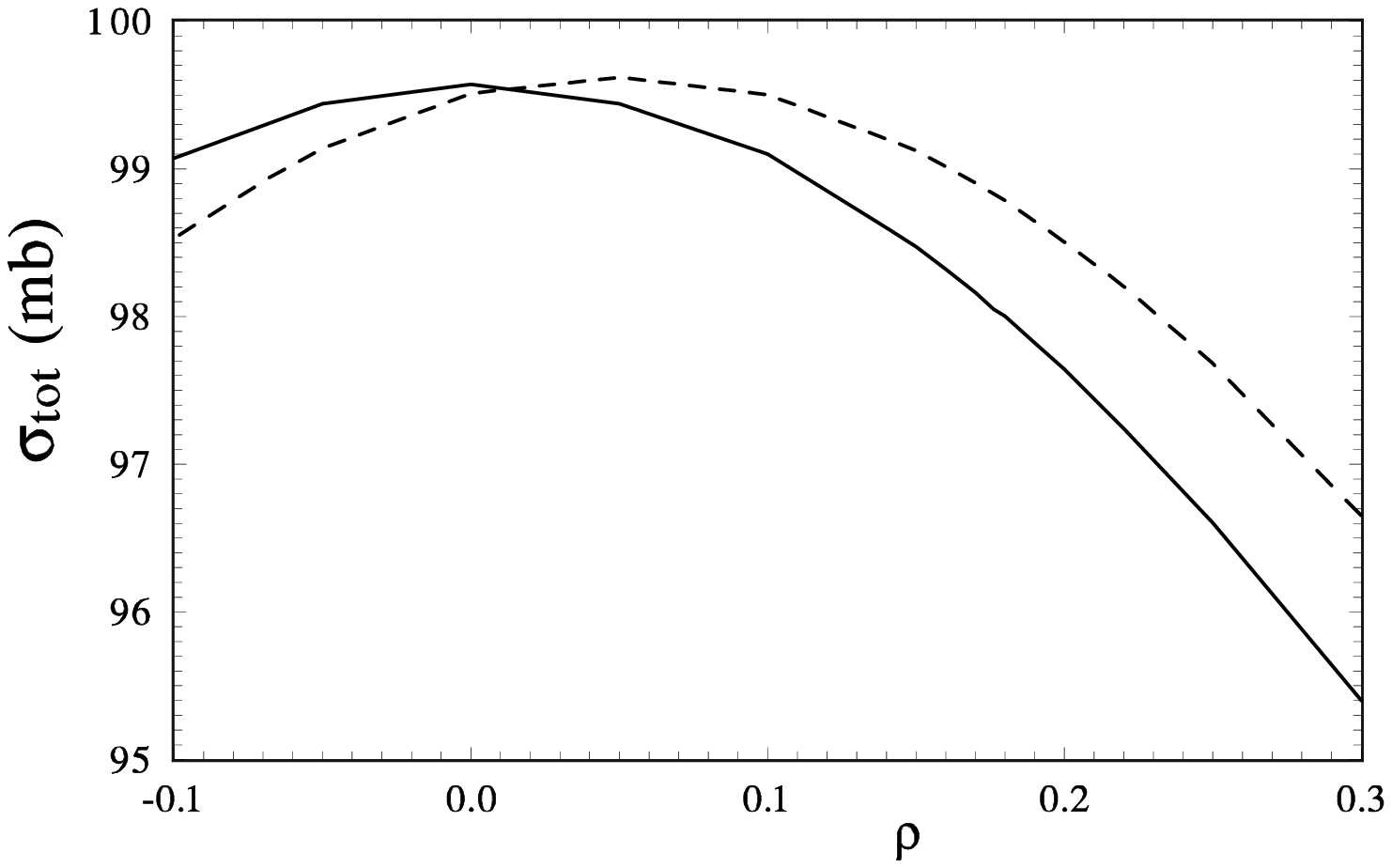}       %ap541L
\includegraphics[width=0.5\textwidth] {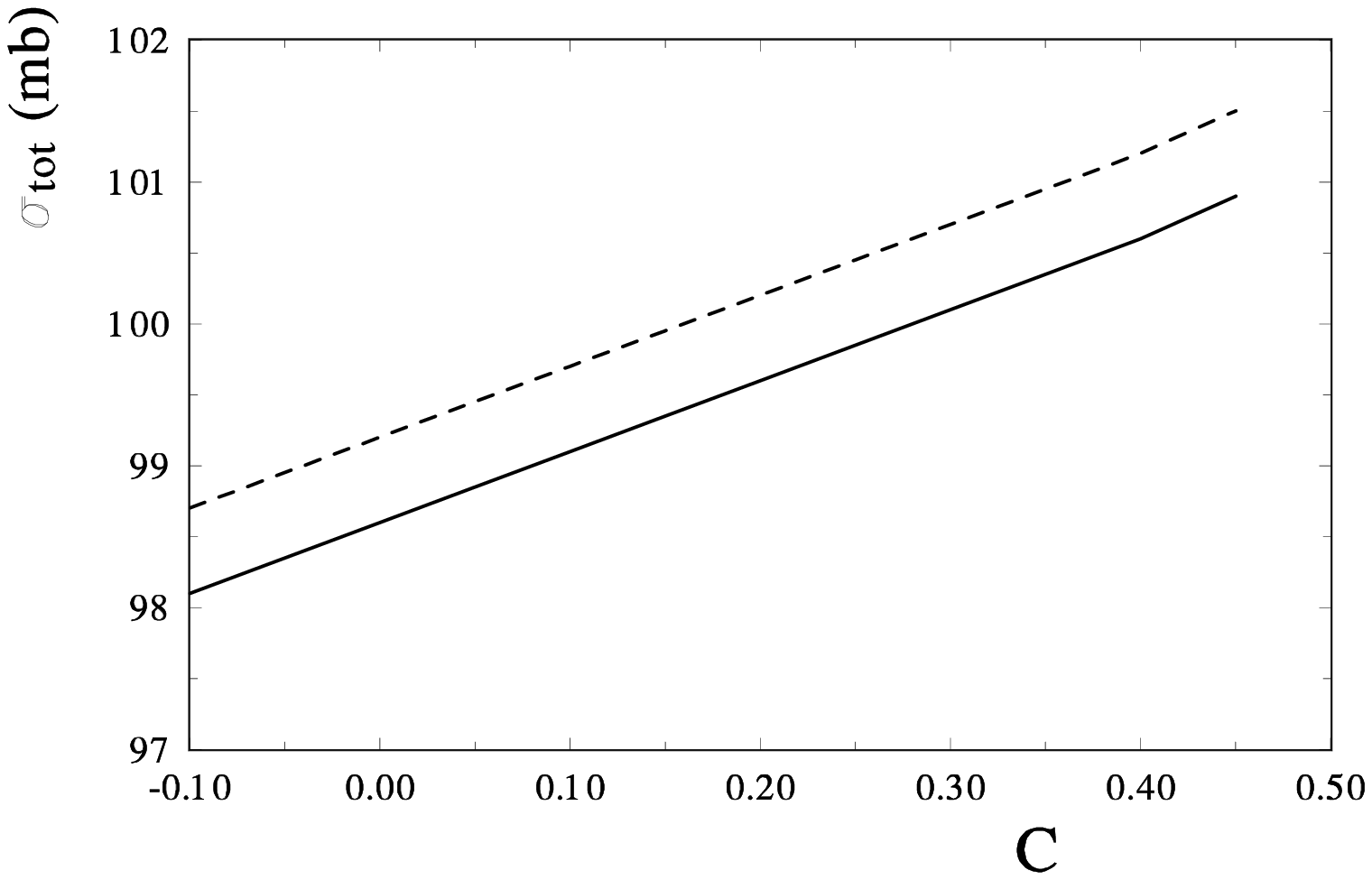}       %ap541L
\caption{ Size of $\sigma_{tot}$ (left)  as a function of $\rho$   and  (right)  $C$
(hard line - without electromagnetic interaction
and dashed line with electromagnetic interaction).
}\label{Fig:1}
\end{figure}

%\newpage

  Now let us make the fit with the fixed size of $\sigma_{tot}= 98.4 \ $mb (see Table 6 for a wide interval of $t$;
  and  Table 7, for a narrow interval of $t$).
  In this case, if we take $C=0$,  $\chi^2$ increases essentially.  The size of $\rho$ is determined well with the
  fixed $C$ and with the free $C$. Again, as in the first case, we obtain the negative size of $\rho$  near value
  $-0.1 \ $ for  both the cases and in the cases of the short and long intervals of $t$.
  The size of the coefficient $C$ is determined badly in both the cases and has a negative sign, but its size is less
  than the size of $C$, presented in Tables 2-5.

     In Fig. 2,  the dependence of the size of $\sigma_{tot}$ as a function of $\rho$ (left picture) and as a function of
     slope $C$ (right picture) in the case without and with the contributions of the Coulomb and Coulomb-hadron
     interference terms  is shown. The inclusion of the Coulomb dependence terms leads to an increase of $\sigma_{tot}$
     at large $\rho$ and decrease in the case of small and negative $\rho$. Contrary, the dependence of the size of $\sigma_{tot}$
     on the coefficient $C$ is linear. % So $C=0$ leads to the minimum size of  $\sigma_{tot}$.
     From Tables 2-5 it is clear that the negative size of $C$ can be only with large negative $\rho$.
     The minimum  $\sum_{i=1}^{N} \chi^{2}_{i}$  was obtained with large positive $C$. Again, we can see that the inclusion of the Coulomb
     dependence terms leads to an increase of the size of  $\sigma_{tot}$.

   In our opinion, this analysis shows the problems when we try to measure $\sigma_{tot}(s)$ and $\rho(s,t)$ in the
   different experiments. Hence, to avoid  such a situation, we need to determine  $\sigma_{tot}(s)$ and $\rho(s,t)$
   simultaneously in one experiment.

%Conclusion:
%  It is need measure
%  $ \sigma_{tot}$ and $\rho$ simultaneously.

%  %==================Figs.3 =============================
%\label{sec:figures}
%\begin{figure}[htb]
%%\centerline{\includegraphics[width=0.35\textwidth]{smith_joe_eds09_fig1.eps}}
%\begin{center}
%\includegraphics[width=0.5\textwidth] {totnc0cf.ps}       %ap541L
%%\includegraphics[width=0.5\textwidth] {rcexth.ps}       %ap541L
%%\includegraphics[width=0.5\textwidth] {rtm1shei.ps}
%\end{center}
%\caption{ Size of $\sigma_{tot}$ over $n$ in two variants:(hard line) - with free slope $C$
%and (dashed line) with $C=0$. \hspace{10.cm}
%}\label{Fig:3}
%\end{figure}

%\newpage

%\section{Additional normalization of the differential cross sections. % by $n$
%, which is reflect the systematical errors, and [$\sigma_{tot}= 98.4$ mb is fixed]
%}

\section{Regge representation}

   In the previous analysis  the simplest phenomenological model with many assumptions (for example,
   the equality of the slopes of the imaginary and real parts  and others)  was used.
   Now let us take also the  simple model  but which is based on the Regge representation for the scattering amplitude
   and was used in the Donnachi-Landshoff \cite{DL-hp} model.
   In this case,  the real part is not fitted but is determined by the form of the scattering amplitude.
   We take the hadron form-factor in the standard electromagnetic form. Hence, some part of the $t$ dependence of the
   scattering amplitude is fixed.

The hadron spin non-flip amplitude was chosen in the form
\begin{eqnarray}
 F(s,t)= i h \hat{s}^{\Delta} f_{1}(t)^2 \ \frac{\sigma_{tot}}{4 \pi \ 0.38938} \
  e^{[\alpha_{1}^{\prime} \ t \ + \alpha_{2}^{\prime} \ (\sqrt{4\mu^2-t}-2\mu)] \ Ln(\hat{s}) }
 \label{fhR}
  \end{eqnarray}
with the electromagnetic form factor $f_{1}(t)$ (\ref{emff}) and $\hat{s}=s e^{-i*\pi/2} $; $\mu$ is the pion mass.
We take $ h Re( \hat{s}^{\Delta}) =1$ at $\sqrt{s}=7 \ $TeV.

% Table 2
    \begin{table}[htb]
\label{tab:13}       % Give a unique label
\begin{center}
\begin{tabular}{c|c|c|c|c}
\hline\noalign{\smallskip}
 \multicolumn{5}{c} {$ \Delta=0.1$ \ \  \ $\rho(\sqrt{s} =7  \ $TeV, $ t=0)=0.156 $ } \\ \hline
  %$0.000875 \leq |t| \leq 0.12 \ $GeV$^2$)}   \\
\noalign{\smallskip}\hline\noalign{\smallskip}
%  &  & $ \Delta=0.1$ & $\rho(0)=0.156$ &   \\\hline
 $N$  & $\sum_{i=1}^{N} \chi^{2}_{i}$  & $\alpha_{1}^{\prime}$ & $\alpha_{2}^{\prime}$ & $\sigma_{tot}, mb$  \\\hline
  & & & & \\
 47 & 65.2   &$0.325$  & $0.fix           $ & $ 99.7\pm0.15  $   \\
 47 & 65.1   &$0.328$  & $-0.002\pm0.015   $ & $ 99.6\pm0.2   $   \\
 40 & 58.1   &$0.324$  & $ 0.fix          $ & $ 99.6\pm0.05  $    \\
 40 & 55.8   &$0.276$  & $ 0.037\pm0.02 $ & $ 99.9\pm0.26   $    \\
 47 & 204    &$0.314$  & $ 0.fix          $ & $  98.4 fix    $     \\
 47 & 95.3   &$0.427$  & $-0.075\pm0.008 $ & $  98.4 fix    $    \\
 40 & 154    &$0.312$  & $  0.fix         $ & $  98.4 fix    $    \\
 40 & 90.1   &$0.437$  & $-0.08\pm0.01 $ & $  98.4 fix    $    \\
\noalign{\smallskip}\hline
\end{tabular}
\caption{The basic parameters of the Regge amplitude are determined by fitting experimental data with free $\sigma_{tot}$.}
\end{center}
\end{table}
% Table 2
    \begin{table}[htb]
\label{tab:14}       % Give a unique label
\begin{center}
\begin{tabular}{c|c|c|c|c}
\hline\noalign{\smallskip}
 \multicolumn{5}{c} {$ \Delta=0.08$ \ \  \ $\rho(\sqrt{s} =7  \ $TeV, $ t=0)=0.128 $ } \\ \hline
 \noalign{\smallskip}\hline\noalign{\smallskip}
%  &  & $ \Delta=0.08$ & $\rho(0)=0.128$ &   \\\hline
 $N$  & $\sum_{i=1}^{N} \chi^{2}_{i}$  & $\alpha_{1}^{\prime}$ & $\alpha_{2}^{\prime}$ & $\sigma_{tot}, mb$  \\\hline
  & & & & \\
 47 & 64.4   &$0.325$  & $0.fix           $ & $ 100.0\pm0.1  $   \\
 47 & 64.3   &$0.338$  & $0.008\pm0.003   $ & $ 99.9\pm0.1   $   \\
 40 & 56.4   &$0.323$  & $ 0.fix          $ & $ 99.9\pm0.12  $    \\
 40 & 55.3   &$0.299$  & $ 0.005\pm0.004 $ & $100.2\pm0.3   $    \\
 47 & 285    &$0.310$  & $ 0.fix          $ & $  98.4 fix    $     \\
 47 & 106.7   &$0.455$  & $-0.096\pm0.008 $ & $  98.4 fix    $    \\
 40 & 211    &$0.307$  & $  0.fix         $ & $  98.4 fix    $    \\
 40 & 99.6   &$0.473$  & $-0.108\pm0.011 $ & $  98.4 fix    $    \\
\noalign{\smallskip}\hline
\end{tabular}
\caption{The basic parameters of the model are determined by fitting experimental data with free $\sigma_{tot}$.}
\end{center}
\end{table}

We examined two  sizes of the intercept $ \alpha(0)=1+\Delta$  with
   a) $\Delta=0.1$ (Table 8) and b) $\Delta=0.08$ (Table 9). These sizes of the intercept lead
   to  $\rho(0)=0.156$ and $\rho(0)=0.128$ at $ \sqrt{s} = 7 \ $TeV, respectively.
%   The size of  $\alpha_{1}^{\prime}$ has a small change in the fitting procedures in the region $0.31 \div 0.34$.
   The   $\sum_{i=1}^{N} \chi^{2}_{i}$ is slightly above the minimal  $\sum_{i=1}^{N} \chi^{2}_{i}$
    in the pure phenomenological cases if we do not fix
   $\sigma_{tot}$. Contrary, if we take the size of $\sigma_{tot} = 98.4 \ $mb, as obtained by
    the TOTEM Collaboration,  $\chi^2$ increases essentially, especially we do not include in the fitting
   procedure the additional slope $\alpha_{2}^{\prime}$. This coefficient is determined if we examine the short interval of
   momentum transfer with the number of experimental points $40$.

 %  Comparing the results in  Tables 8 and 9 we see that  $\sum_{i=1}^{N} \chi^{2}_{i}$ is slightly less in the second case.
 %  However, the obtained sizes of the total cross sections are practically the same.
 %  Hence, the taken interval of the intercept and correspondingly $\rho$ lead to the small difference
 %  in the fitting sizes of $\sigma_{tot}$.%

%    Comparing the results of the Regge approaches with the final results of the pure phenomenological model
%    % (Table 10)
%    we see that the maximal $\sigma_{tot}$ are the same, near $100.9 \ $mb, but the minimal sizes are
%    less in the case of the pure phenomenological model, near $97.2 \ $mb.

  Comparing the results in  Tables 8 and 9 we see that  $\sum_{i=1}^{N} \chi^{2}_{i}$ is slightly less in the second case.
   However, the obtained sizes of the total cross sections are practically the same.
   Hence, the taken interval of the intercept and correspondingly $\rho$ lead to the small difference
   in the fitting sizes of $\sigma_{tot}$. Such representation shows that the slope has some complicated form and
   it is necessary to be taken into account in the pure phenomenological analysis.

    Comparing the results of the Regge approaches with the  results of the pure phenomenological model
    we see that the maximal $\sigma_{tot}$ are the same, near $100.9 \ $mb, but the minimal sizes are
    less in the case of the pure phenomenological model, near $97.18 \ $mb. In last case the size of $\rho$ is near zero.
      Hence the value of $\rho \sim 0.1 \div 0.15$ leads a larger   $\sigma_{tot}$ than obtained
      by the TOTEM Collaboration.

 \section{The size of $\rho$}

    Let us check up our assumptions about the size and momentum transfer dependence of the real part of the scattering
    amplitude. We can use the method which was proposed and explored in \cite{Sel-Bl95,Sel-Hel03,Sel-AW,Sel-PL05},
    and introduce the value
 \begin{eqnarray}
    \Delta_{R}^{th}(s,t)=(ReF_C(t)+ReF_h(s,t))^2 \geq 0.
    \label{Dr}
\end{eqnarray}

From the eq.(9)
% (\ref{refh}), one can obtain an equation for $\Delta_R(s,t)$  for every experimental point $i$:
 \begin{eqnarray}
  \Delta_{R}^{exp}(s,t_{i}) = [\frac{d\sigma}{dt_{i}}|_{exp.}/n - k \pi *(ImF_c(t_{i}) + ImF_h(t_{i}))^2]/(k \pi).
%   \\ \nonumber
\label{Drds}
  \end{eqnarray}
  where $n$ is the additional normalization coefficient which is reflect the errors in the Luminosity determination and other possible systematical errors.
    For the proton-proton high energy scattering the real part of the hadron scattering amplitude is positive at small momentum transfer,
    and Coulomb amplitude is  negative and exceeds the size of the hadronic part of the amplitude
    at $t \rightarrow 0$, but has a large slope. Hence,  $\Delta_R(t)$
    will have the minimum at some value  $t$ and then a wide maximum.
    Comparing $\Delta_{R}^{th}(s,t)$ with $ \Delta_{R}^{exp}(s,t_{i})$ gives the accuracy of the experiment and of the theoretical
    model assumptions.

   Let us take the parameters obtained the TOTEM Collaboration $\sigma_{tot}=98.6$ mb, $B=19.9 $ GeV$^{-2}$, $n=1$,
    $\rho(0)=0.141$ and calculate the value  $\Delta_{R}^{th}(s,t)$, eq.({\ref{Dr}). The result is shown in Fig.3 by the hard line.
    Now let us take these parameters for the imaginary part of the scattering amplitude and calculate $\Delta_{R}^{exp}(s,t_{i})$  using eq.({\ref{Drds}). The triangles  in Fig.3 present these calculations. Obviously, the first and second calculations are very far from each other. If we take the real part with the parameters
    $\sigma_{tot}=96.4$ mb, $B=19.9 \ $ GeV$^{-2}$, $\rho=0.1$ and calculate $\Delta_{R}^{th}(s,t)$    (short  dashed line in Fig.3)
    the position of the minimum moves to higher $t$, but the difference remains large.

     %==================Figs.4 =============================
\label{sec:figures}
\begin{figure}[htb]
\vspace{-1.5cm}
\begin{center}
%\centerline{\includegraphics[width=0.35\textwidth]{smith_joe_eds09_fig1.eps}}
%\includegraphics[width=0.5\textwidth] {totnc0cf.ps}
\includegraphics[width=0.7\textwidth] {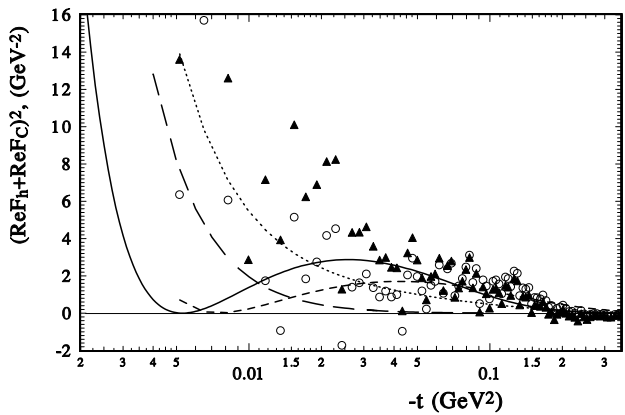}
\end{center}
\caption{  $\Delta_{R}^{exp}(t_{i})$ eq.(\ref{Drds}): triangles - calculations with the TOTEM parameters;  circles - calculations
with $\sigma_{tot}=96.4$ mb, $B=20.3 \ $ GeV$^{-2}$,  $n=1.08$, $C=-0.05$; hard line - eq.(\ref{Dr}) the real part with
TOTEM parameters and $\rho=0.14$; short dashed line -  $\Delta_{R}^{th}(s,t)$     eq.(\ref{Dr}) with $\sigma_{tot}=96.4$ mb, $B=19.9 \ $ GeV$^{-2}$
and $\rho=0.1$; long dashed line - eq.(\ref{Dr}) with $\rho=0$; points line - eq.(\ref{Dr}) with $\sigma_{tot}=96.4$ mb,
 $B=19.9 \ $GeV$^{-2}$ and $\rho=-0.05$.
}
\label{Fig:4}
\end{figure}
%\newpage

 If we take some other parameters for the imaginary part $\sigma_{tot}=96.4$ mb, $B=20.3 $ GeV$^{-2}$,
        $n=1.08$, $C=-0.05$ GeV$^{-1}$, we obtain from  eq.({\ref{Drds}) the results  for $\Delta_{R}^{exp}(t_{i})$,
          which is shown in Fig.3 by circles.
    The $\Delta_{R}^{th}(s,t)$ with only the Coulomb amplitude ($\rho=0$)  is shown in Fig.3
    by the long dashed line, and the dotted line presents the calculation with $\rho=-0.05$. We can see that only
    in the last case the difference between the calculations by  eqs.({\ref{Drds}) and ({\ref{Dr})
    is not large. That is why  our fitting procedure in most part requires the size of $\rho$ near zero or negative.
    But these parameters for the imaginary  and real parts of the scattering amplitude are far from the parameters
    obtained by the TOTEM Collaboration. This situation is unclean. Maybe, there is some problem with the
    normalization of the separate parts of the experimental data, or there exists some additional (probably oscillation)
    term (see \cite{Osc}) which changes the form of the imaginary part.

 \section{Conclusion}

   The analysis of the new experimental data obtained by the LHC TOTEM Collaboration \cite{TOTEM-11a,TOTEM-11}
   shows that there are some additional specific moments which are to be  taken into account in determining the size of the
   total cross sections.
   % First, we cannot neglect the Coulomb-hadron interference term.
    We found that it is necessary to take into account  the electromagnetic interactions in the analysis of the experimental data.
    In this case, the impact of the size of $\rho$ on the  determination of $\sigma_{tot}$ increases in comparison with the standard
   factor $1/\sqrt{1+\rho^2}$.
   %Second,
   The deviation of the hadron part of the scattering amplitude at small momentum transfer
   from the standard exponential form
   can be taken into account by the additional part of the slope proportional to $q$. This impact will increase
   when the new experimental data will be obtained for smaller $t$.
%   Third, the errors of the Luminosity can be taken into account by an additional normalization coefficient.
%   Fourth,
  It is needed to check out the obtained, during the fitting procedure,  real part of the hadron amplitude
   by using eq.({\ref{Drds}).
    Now our calculations show the inconsistency of the size of $\rho=0.14$ with the parameters of the scattering amplitude
    obtained  by the TOTEM Collaboration.
    Maybe, it is necessary to explore in more detail
     the non-exponential behavior of the real and imaginary part  s
   and a possibility of the presence of some oscillation term.

%      We have not discussed here the Luminosity independent method. There are some additional problems
%   (for example, in any case we need to obtain $dN_{hadron - elastic}/dt$ at $t=0$). Hence, all the previous problems
%   are present in this method excluding the additional normalization coefficient. However,  the problems of  determination
%   of inelastic processes are  added.

     Finally, we should note that the best way to decrease the impact of the different assumptions, which are examined
     in the phenomenological model, consists in the  determination of
    the sizes of $\sigma_{tot}$ and $\rho(s,t)$ simultaneously in one experiment.

%\newpage

%\vspace*{0.5cm}
%{\bf Acknowledgments}

%We wish to acknowledge
%This work was supported by

\end{document}